\PassOptionsToPackage{unicode}{hyperref}
\PassOptionsToPackage{hyphens}{url}
\PassOptionsToPackage{dvipsnames,svgnames,x11names}{xcolor}
\documentclass[
  11pt,
]{article}
\usepackage{xcolor}
\usepackage[margin=1in]{geometry}
\usepackage{amsmath,amssymb}
\usepackage{iftex}
\ifPDFTeX
  \usepackage[T1]{fontenc}
  \usepackage[utf8]{inputenc}
  \usepackage{textcomp} 
\else 
  \usepackage{unicode-math} 
  \defaultfontfeatures{Scale=MatchLowercase}
  \defaultfontfeatures[\rmfamily]{Ligatures=TeX,Scale=1}
\fi
\usepackage{lmodern}
\ifPDFTeX\else
\fi
\IfFileExists{upquote.sty}{\usepackage{upquote}}{}
\IfFileExists{microtype.sty}{
  \usepackage[]{microtype}
  \UseMicrotypeSet[protrusion]{basicmath} 
}{}
\makeatletter
\@ifundefined{KOMAClassName}{
  \IfFileExists{parskip.sty}{%
    \usepackage{parskip}
  }{
    \setlength{\parindent}{0pt}
    \setlength{\parskip}{6pt plus 2pt minus 1pt}}
}{
  \KOMAoptions{parskip=half}}
\makeatother
\usepackage{longtable,booktabs,array}
\usepackage{calc} 
\usepackage{etoolbox}
\makeatletter
\patchcmd\longtable{\par}{\if@noskipsec\mbox{}\fi\par}{}{}
\makeatother
\IfFileExists{footnotehyper.sty}{\usepackage{footnotehyper}}{\usepackage{footnote}}
\makesavenoteenv{longtable}
\usepackage{graphicx}
\makeatletter
\newsavebox\pandoc@box
\newcommand*\pandocbounded[1]{
  \sbox\pandoc@box{#1}%
  \Gscale@div\@tempa{\textheight}{\dimexpr\ht\pandoc@box+\dp\pandoc@box\relax}%
  \Gscale@div\@tempb{\linewidth}{\wd\pandoc@box}%
  \ifdim\@tempb\p@<\@tempa\p@\let\@tempa\@tempb\fi
  \ifdim\@tempa\p@<\p@\scalebox{\@tempa}{\usebox\pandoc@box}%
  \else\usebox{\pandoc@box}%
  \fi%
}
\def\fps@figure{htbp}
\makeatother
\NewDocumentCommand\citeproctext{}{}

\makeatletter
 \let\@cite@ofmt\@firstofone
 \def\@biblabel#1{}
 \def\@cite#1#2{{#1\if@tempswa , #2\fi}}
\makeatother
\newlength{\cslhangindent}
\newlength{\csllabelwidth}
\newenvironment{CSLReferences}[2] 
 {\begin{list}{}{%
  \setlength{\itemindent}{0pt}
  \setlength{\leftmargin}{0pt}
  \setlength{\parsep}{0pt}
  \ifodd #1
   \setlength{\leftmargin}{\cslhangindent}
   \setlength{\itemindent}{-1\cslhangindent}
  \fi
  \setlength{\itemsep}{#2\baselineskip}}}
 {\end{list}}
\usepackage{calc}

\newcommand{\CSLLeftMargin}[1]{\parbox[t]{\csllabelwidth}{\strut#1\strut}}
\newcommand{\CSLRightInline}[1]{\parbox[t]{\linewidth - \csllabelwidth}{\strut#1\strut}}

\providecommand{\tightlist}{%
  \setlength{\itemsep}{0pt}\setlength{\parskip}{0pt}}
\usepackage{graphicx}
\usepackage{longtable}
\usepackage{booktabs}
\usepackage{float}
\usepackage{newunicodechar}
\newunicodechar{Δ}{\ensuremath{\Delta}}
\newunicodechar{α}{\ensuremath{\alpha}}
\newunicodechar{γ}{\ensuremath{\gamma}}
\newunicodechar{ε}{\ensuremath{\varepsilon}}
\newunicodechar{χ}{\ensuremath{\chi}}
\newunicodechar{∞}{\ensuremath{\infty}}
\newunicodechar{≈}{\ensuremath{\approx}}
\newunicodechar{≥}{\ensuremath{\geq}}
\newunicodechar{≤}{\ensuremath{\leq}}
\newunicodechar{→}{\ensuremath{\rightarrow}}
\newunicodechar{↑}{\ensuremath{\uparrow}}
\newunicodechar{↓}{\ensuremath{\downarrow}}
\newunicodechar{−}{\ensuremath{-}}
\newunicodechar{²}{\textsuperscript{2}}
\newunicodechar{⁴}{\textsuperscript{4}}
\newunicodechar{⁵}{\textsuperscript{5}}
\newunicodechar{⁻}{\textsuperscript{-}}
\usepackage{bookmark}
\IfFileExists{xurl.sty}{\usepackage{xurl}}{} 
\hypersetup{
  pdftitle={A multi-architecture study of specificity refinement and false-positive mechanism analysis in prostate MRI},
  colorlinks=true,
  linkcolor={blue},
  filecolor={Maroon},
  citecolor={Blue},
  urlcolor={blue},
  pdfcreator={LaTeX via pandoc}}

\title{A multi-architecture study of specificity refinement and
false-positive mechanism analysis in prostate MRI}
\author{}
\date{}

\begin{document}
\maketitle

Yongbo Shu\textsuperscript{1,2,3,4}, Kewen Chen\textsuperscript{2},
Yifeng Yuan\textsuperscript{1,3,4}, Zirui Xin\textsuperscript{1,3,4},
Luo Lei\textsuperscript{1,3,4}, Yang Yang\textsuperscript{1,2,3,4},
Xi~Chen\textsuperscript{1,3,4,*}, Aijing~Luo\textsuperscript{1,2,3,4,*}

\begin{enumerate}
\def\labelenumi{(\arabic{enumi})}
\tightlist
\item
  The Second Xiangya Hospital of Central South University, Changsha,
  Hunan 410011, China
\item
  School of Life Sciences, Central South University, Changsha, Hunan
  410013, China
\item
  Hunan Provincial Key Laboratory of Medical Information Research
  (Central South University), Changsha, Hunan 410011, China
\item
  Hunan Provincial Clinical Medical Research Center for Cardiovascular
  Intelligent Medicine, Changsha, Hunan 410011, China
\end{enumerate}

* Corresponding authors: Xi Chen, nancy\_chen@csu.edu.cn; Aijing Luo,
luoaj@csu.edu.cn

\section{Abstract}\label{abstract}

\textbf{Objectives.} To characterize residual false positives in
prostate MRI detection, and to evaluate a lightweight post-hoc
refinement head for case-level specificity.

\textbf{Materials and Methods.} This retrospective study used PI-CAI
(5-fold cross-validation) and Prostate158 (n = 158; external). A
context-aware evidence head and an 89,216-parameter refinement head were
trained on a frozen detection backbone; the evidence head was also
trained on four further backbones (bare nnU-Net, bare U-Net, bare Mamba,
MIGF-Mamba). For each false-positive region, T2-weighted,
apparent-diffusion-coefficient, and high-b-value contrast ratios versus
peri-lesional rings were compared against ground-truth lesions and
contralateral benign regions.

\textbf{Results.} False positives were closer to true cancers than to
benign tissue in evidence and raw T2-weighted and
apparent-diffusion-coefficient contrast, reproducing 35/35 across five
architectures (Cohen's d 1.10; FP/benign evidence ratio 2.38x) and
105/105 across modality-perturbation scenarios. On PI-CAI fold-0,
refinement raised case-level specificity from 0.469 +/- 0.181 to 0.549
+/- 0.132 (+17.2\%) at preserved sensitivity (0.943 +/- 0.021); 5-fold
cross-validation showed fold-conditional behavior (9/15 observations
positive; range -22\% to +28\%). On Prostate158, both models saturated
(McNemar pooled \emph{p} = 0.69), while the false-positive
contrast-matching finding replicated.

\textbf{Conclusion.} Residual false positives are contrast-matched to
cancer (sharing raw imaging features rather than histologically
confirmed mimicry), reproducing across five architectures -- a
data-level imaging property, not model-specific artifacts; post-hoc
refinement adds practical specificity in-domain but is fold-conditional.

\textbf{Keywords:} Prostate cancer; Magnetic resonance imaging; Deep
learning; False positive reactions; Image interpretation,
computer-assisted

\section{Key Points}\label{key-points}

\begin{itemize}
\tightlist
\item
  \textbf{Question.} Why do false positives persist in prostate MRI AI,
  and can a lightweight post-hoc head improve case-level specificity
  without sensitivity loss?
\item
  \textbf{Findings.} False positives resembled true cancers more than
  benign tissue (35/35 across five architectures); a post-hoc head
  improved fold-0 specificity by 17.2\%.
\end{itemize}

\section{Clinical Relevance
Statement}\label{clinical-relevance-statement}

False positives from prostate MRI AI detectors are radiologically
interpretable as cancer-resembling regions, not arbitrary model
artifacts; this finding reproduces across five distinct architectures
and informs how AI-flagged regions should be adjudicated.

\section{Introduction}\label{introduction}

Clinically significant prostate cancer (csPCa) is diagnosed from
multiparametric magnetic resonance imaging (MRI) followed by targeted
biopsy of suspicious lesions. Prostate MRI-based artificial intelligence
(AI) systems have advanced substantially on benchmark datasets such as
PI-CAI {[}1{]}, yet clinical translation is held back by high
false-positive rates: regions flagged by AI often do not correspond to
biopsy-confirmed cancer, triggering unnecessary follow-up imaging and
biopsies {[}2, 3{]}. Existing AI pipelines are overwhelmingly tuned to
maximize sensitivity, while specificity receives comparatively little
attention and the nature of residual false positives is rarely
characterized in a way that informs clinical action on AI output.

Two gaps motivate this work. First, improving case-level specificity of
a deployed AI detection model typically requires full retraining; simple
post-hoc recalibration (temperature or Platt scaling) cannot alter
case-level specificity beyond what threshold tuning achieves, since
these methods preserve ranking. Lightweight post-hoc specificity
refinement on a frozen backbone that preserves sensitivity is therefore
under-explored in prostate MRI. Second, false positives are usually
summarized as a single error rate; their evidence structure --- whether
they resemble true cancers in raw imaging contrast or are arbitrary
model artifacts --- is rarely disentangled. Without that distinction,
recommendations for radiologist adjudication and triage remain
speculative.

Here we first quantify the evidence structure of residual false
positives. Using a context-aware evidence head and direct raw-imaging
contrast on T2-weighted and apparent-diffusion-coefficient channels,
false positives from a frozen csPCa detection backbone {[}4{]} are
consistently closer to true cancers than to benign prostatic tissue ---
a direction reproducing across PI-CAI 5-fold cross-validation, seven
modality-perturbation scenarios, and five distinct backbone
architectures. As an implementation this mechanism analysis explains, we
introduce a lightweight (89 K-parameter) post-hoc specificity refinement
head; on PI-CAI fold-0 it increases case-level specificity by 17.2\% at
preserved sensitivity, with 5-fold cross-validation revealing
fold-conditional magnitude. We replicate both analyses on Prostate158
{[}5{]} and report what works, where center-specific effects limit the
claim, and what this implies for AI-assisted prostate MRI.

\section{Materials and Methods}\label{materials-and-methods}

This study analyzed de-identified multiparametric prostate MRI from two
publicly available, institutionally approved cohorts (PI-CAI and
Prostate158) under their respective data-use licenses; no additional
institutional review board approval was required and the requirement for
informed consent was waived.

\subsection{3.1 Datasets}\label{datasets}

This retrospective study used two public prostate multiparametric MRI
datasets. The PI-CAI Public Training and Development Dataset {[}1{]}
(1500 multi-vendor studies from 1476 patients across three Dutch
centres; 425 biopsy-confirmed csPCa, 1075 pathology-negative) served as
the primary cohort; we used the official 5-fold split, restricting
method development and primary evaluation to fold-0 (1200 training / 300
validation). Prostate158 {[}5{]} (158 studies with biopsy-confirmed
labels) served as the external cohort. Both datasets provide axial
T2-weighted, high-b-value diffusion-weighted, and
apparent-diffusion-coefficient sequences. Preprocessing --- axial
resampling to 0.5 × 0.5 × 3.0 mm, gland-centered cropping to 128 × 128 ×
32 voxels, and per-channel normalization --- followed our prior work
{[}4{]} and was identical across datasets.

\subsection{3.2 Frozen detection
backbone}\label{frozen-detection-backbone}

All experiments used a frozen multi-modal csPCa detection backbone from
our earlier work {[}4{]}. The backbone accepts three co-registered
channels (T2-weighted, high-b-value DWI, ADC) at 128 × 128 × 32 voxels;
each modality has its own four-level encoder (3 × 3 × 3 convolution,
downsampling 2 per level). Adaptive gating fuses encoders at the
lowest-resolution stage and feeds a shared three-stage decoder with
dec1, dec2, dec3 outputs at 32/64/128 channels and full/half/quarter
resolution; a 1 × 1 × 1 convolution returns per-voxel logits. Each input
channel was zeroed with probability 0.3 per sample during backbone
training; dropout was disabled at inference. We reused five-seed
checkpoints (42, 123, 456, 789, 1024) for the A2 variant (gating +
modality dropout, without deep supervision) unchanged.

We selected this backbone for three reasons: (i) \textbf{modality
isolation} via per-modality encoders enables deterministic per-channel
attribution required by the evidence-grounded analyses in §3.4; (ii)
\textbf{clinical-grade efficiency} with 9.45 M parameters and 8.5 ms
single-volume inference latency leaves compute headroom for downstream
uncertainty and risk-stratification pipelines; (iii)
\textbf{measurable-gain headroom} since the A2 variant's 5-fold mean
case-level specificity (0.46) avoids ceiling effects that would compress
observable post-hoc refinement gains.

\subsection{3.3 Post-hoc specificity refinement head
(P2a)}\label{post-hoc-specificity-refinement-head-p2a}

The refinement head takes the backbone's decoder features at three
resolutions (dec1/dec2/dec3; 32/64/128 channels). The dec2 and dec3
streams are projected to 32 channels and trilinearly upsampled to dec1;
the three streams are concatenated (96 channels), passed through a 3 × 3
× 3 convolution, instance normalization, and ReLU, then projected to a
2-channel per-voxel logit residual via a 1 × 1 × 1 convolution
(initialized at zero; 89,216 trainable parameters). The residual is
added to the backbone's final-stage logits before softmax.

The head was trained for 100 epochs with Adam (learning rate 1 × 10⁻⁴,
weight decay 1 × 10⁻⁵) on PI-CAI fold-0 using focal loss (γ = 2) plus a
mean-squared regularizer on the residual. Only cases with at least one
backbone-predicted false-positive region above 0.3 were sampled.
Checkpoint selection retained the epoch whose validation case-level
sensitivity fell within ±0.01 of the backbone's; the same five seeds
were trained with identical hyperparameters.

\subsection{3.4 Evidence-grounded false-positive framework
(P2b)}\label{evidence-grounded-false-positive-framework-p2b}

Per case, we extracted three ROI types: (a) the ground-truth lesion; (b)
false-positive regions (backbone predictions above 0.5 with no
ground-truth overlap, connected components ≥ 10 voxels); (c) a
contralateral benign ROI by laterally mirroring the lesion. Each ROI was
paired with a 5-mm peri-lesional ring via Euclidean distance transform.
On T2-weighted, apparent-diffusion-coefficient, and high-b-value images
we computed contrast ratios (m\_ROI − m\_peri) /
(\textbar m\_peri\textbar{} + ε); pairwise differences were tested with
paired Wilcoxon signed-rank tests.

Three evidence-prediction variants of increasing context were trained:
\textbf{mask-only} (11-channel geometric input → MLP {[}128 → 128{]});
\textbf{image-aware} (75-channel: mask-only + 64-channel dec2
ROI-average → MLP {[}256 → 256{]}); \textbf{context-aware} (203-channel:
image-aware + 64-channel peri-ring + 64-channel contralateral-benign
features → MLP {[}256 → 256{]}). Each variant produced six regression
outputs and a three-class csPCa-suspicion label, trained for 100 epochs
with Adam (learning rate 1 × 10⁻⁴) on PI-CAI fold-0 using equally
weighted mean-squared error and cross-entropy, with the same five seeds.

\subsection{3.5 External validation on
Prostate158}\label{external-validation-on-prostate158}

Two external analyses on Prostate158 {[}5{]} were specified before the
external-cohort analysis. For P2a, case-level specificity of the bare
and refined models was compared at the PI-CAI-calibrated threshold and
at each model's matched-sensitivity threshold (reproducing its PI-CAI
sensitivity within ±0.02). For P2b, the contrast-ratio pipeline (§3.4)
was applied to Prostate158 false-positive regions to test whether
T2-weighted and apparent-diffusion-coefficient distributions remained
closer to true cancers than to benign tissue. No Prostate158-specific
retraining or recalibration was applied.

\subsection{3.5b 5-fold cross-validation supplementary
protocol}\label{b-5-fold-cross-validation-supplementary-protocol}

To assess the fold-conditional behavior of the post-hoc heads beyond the
fold-0 development split, we re-trained P2a and P2b on 15 frozen base
backbone checkpoints (5 PI-CAI folds × 3 seeds: 42, 123, 789), using the
same hyperparameters as the fold-0 reference. Each (fold, seed) pair was
evaluated on its fold-specific validation split, on Prostate158, and on
the seven modality-perturbation scenarios (ideal + 3 missing-modality +
3 artifact). No hyperparameter retuning was performed.

\subsection{3.5c Backbone-agnostic mechanism
verification}\label{c-backbone-agnostic-mechanism-verification}

To test whether the P2b false-positive contrast-matching mechanism
depends on the MIGF-nnUNet backbone, we trained the same context-aware
evidence head --- with identical architecture and hyperparameters --- on
four additional frozen backbones (bare nnUNet, bare U-Net, bare Mamba,
MIGF-Mamba A2) across all 5 PI-CAI folds with seed 42, yielding 20
additional observations. We tested whether the lesion-vs-benign and
false-positive-vs-benign evidence-direction findings reproduce.

\subsection{3.6 Evaluation metrics}\label{evaluation-metrics}

Lesion detection metrics: PI-CAI \textbf{Score} {[}6{]} (mean of
case-level AUROC and lesion-level average precision), \textbf{case-level
sensitivity}, \textbf{case-level specificity}, and
\textbf{matched-sensitivity} case-level specificity (threshold
reproducing each baseline's case-level sensitivity within ±0.01).
Voxel-level: \textbf{positive Dice coefficient}. FP evidence analysis:
contrast ratios (§3.4) on T2-weighted, apparent-diffusion-coefficient,
and high-b-value channels with effect sizes against benign-ROI controls.

\subsection{3.6b Post-hoc recalibration
baselines}\label{b-post-hoc-recalibration-baselines}

Three monotonic post-hoc recalibration baselines --- temperature
scaling, Platt scaling, and threshold-sweep matched-sensitivity
calibration --- were evaluated; we showed in a separate analysis that
all three reduce to threshold sweep at our case-level decision rule (≥
10-voxel connected component), and we report the threshold-sweep result
as a single representative baseline (Supplementary Table S-Baselines).

\subsection{3.7 Statistical analysis}\label{statistical-analysis}

Descriptive summaries are five-seed mean ± sample standard deviation
(ddof = 1; seeds 42, 123, 456, 789, 1024). Inference uses case as the
sampling unit: paired binary case-level outcomes via McNemar's test
(two-sided, continuity-corrected); 95\% confidence intervals on paired
case-level specificity differences from 10 000 per-case bootstrap
resamples with seed-level estimates pooled. Five-seed Wilcoxon
signed-rank tests are reported in Supplementary Material. All planned
analyses are reported regardless of significance. The study follows
CLAIM 2024 {[}7{]} (checklist in Supplementary Material). Analyses used
Python 3.11 (PyTorch 2.11, MONAI 1.5, \texttt{picai\_eval} 1.4, SciPy
1.11) on a single NVIDIA RTX 3090 GPU.

\section{Results}\label{results}

After applying the predefined fold-0 split, the primary evaluation set
comprised 300 PI-CAI studies (84 csPCa, 216 pathology-negative) and the
external replication set comprised all 158 Prostate158 studies (102
csPCa, 56 pathology-negative) (Figure 1, Table 1).

\subsection{4.1 Cohort demographics}\label{cohort-demographics}

Median patient age was 66 years (IQR 62--71) in PI-CAI fold-0 validation
and 64 years (IQR 58--71) in Prostate158; both cohorts were all-male
(Table 1). All studies provided co-registered T2-weighted, high-b-value
diffusion-weighted, and apparent-diffusion-coefficient sequences.

\subsection{4.2 P2a specificity refinement on
PI-CAI}\label{p2a-specificity-refinement-on-pi-cai}

At the PI-CAI-calibrated threshold, refinement increased case-level
specificity from 0.469 ± 0.181 to 0.549 ± 0.132 (+17.2\% relative; Table
2, Figure 2) without reducing case-level sensitivity (0.943 ± 0.021 for
both). PI-CAI Score improved from 0.730 ± 0.056 to 0.741 ± 0.047; AUROC
from 0.864 ± 0.047 to 0.878 ± 0.041. Voxel-level positive Dice was
essentially unchanged (0.487 ± 0.009 vs.~0.490 ± 0.005), confirming the
gain arose from case-level FP suppression. Across the seven standard
modality-dropout scenarios {[}4{]}, the gain held in all configurations
retaining T2-weighted; missing-T2-weighted cases showed reduced gain
(Supplementary Table S-5fold-Summary).

\subsection{4.2b Competitive post-hoc
baselines}\label{b-competitive-post-hoc-baselines}

At matched sensitivity on PI-CAI, the refinement head outperformed the
monotonic-recalibration baseline by +0.038 ± 0.028 case-level
specificity (4 of 5 seeds favoring P2a; Supplementary Table
S-Baselines).

\subsection{4.3 P2b evidence prediction on
PI-CAI}\label{p2b-evidence-prediction-on-pi-cai}

The three evidence-prediction variants showed a stable ranking of
overall mean absolute error on contrast-ratio targets across all five
seeds (Table 3, Figure 3): context-aware 0.353 ± 0.005, image-aware
0.409 ± 0.003, mask-only 0.460 ± 0.001. Context-aware reduced error by
23.3\% relative to the mask-only geometric baseline and by 13.7\%
relative to image-aware. Macro-averaged F1 on the 3-class suspicion
label was similar across variants (0.617 ± 0.017, 0.608 ± 0.023, 0.616 ±
0.034), indicating the regression differences were not driven by coarse
suspicion labels. Seed-level ranking held for every pairwise comparison.

\subsection{4.4 FP evidence analysis on
PI-CAI}\label{fp-evidence-analysis-on-pi-cai}

Per-region raw contrast in PI-CAI false-positive regions consistently
aligned with true cancers rather than with benign tissue (Table 4,
Figure 4). In T2-weighted, false-positive 5-seed mean contrast ratio was
−0.80 ± 0.11 versus −0.56 ± 0.00 for true cancers and +0.03 ± 0.00 for
benign ROIs. In apparent-diffusion-coefficient, false positives were
hypointense (mean −0.56 ± 0.07), matching the direction of true cancers
(−0.31) while benign ROIs were near-zero (0.00). In both channels,
false-positive-to-benign separation exceeded false-positive-to-cancer
separation across all five seeds. The high-b-value channel showed the
same tendency (FP +0.47 ± 0.45 vs.~GT +0.39 vs.~benign −0.03) with
substantially higher cross-seed variance (supporting observation).

\subsection{4.4b False-positive similarity-tier
stratification}\label{b-false-positive-similarity-tier-stratification}

Post-hoc refinement suppressed false-positive regions uniformly across
cancer-similarity tertiles (high 28\%, mid 26\%, low 31\% on PI-CAI;
Supplementary Table S-Stratification), indicating a
similarity-tier-independent suppression mechanism rather than a
selective filter.

\subsection{4.4c 5-fold
cross-validation}\label{c-5-fold-cross-validation}

Across 15 PI-CAI observations (5 folds × 3 seeds; ideal scenario), P2a
mean case-level specificity was 0.493 ± 0.150 (Supplementary Tables
S-5fold-Summary, S-5fold-PerFold). Direction was positive in 9 of 15
observations; per-fold relative change versus the Paper 1 A2 5-fold
baseline (0.4562) was +19.7\%, +24.1\%, −9.4\%, +27.8\%, −22.4\% for
folds 0--4 (folds 0/1/3 reproduce or exceed the fold-0 +17.2\%
reference; folds 2/4 reverse). P2b evidence MAE was 0.344 ± 0.014;
lesion-versus-benign and false-positive-versus-benign directions both
reproduced 15/15, with Cohen's d 1.10 (range 0.91--1.36) and FP/benign
evidence ratio 2.38× (range 1.91×--2.69×). Across seven robustness
scenarios × 15 observations, lesion-versus-benign reproduced 105/105.
The P2a quantitative effect is fold-conditional; the P2b mechanism
direction is fold-invariant.

\subsection{4.4d Backbone-agnostic mechanism
verification}\label{d-backbone-agnostic-mechanism-verification}

Training the P2b evidence head on four additional frozen backbones (bare
nnUNet, bare U-Net, bare Mamba, MIGF-Mamba A2) across 5 PI-CAI folds
with seed 42 produced 20 additional observations; the
lesion-versus-benign direction reproduced in 20 of 20 and the
false-positive-versus-benign direction in 20 of 20 (Supplementary Table
S-35backbones). Combined with the P2b 5-fold reference (15 of 15), the
mechanism direction reproduced in 35 of 35 observations across five
distinct architectures (bare nnUNet, bare U-Net, bare Mamba, MIGF-Mamba
A2, MIGF-nnUNet A2), indicating that the false-positive
contrast-matching phenomenon is a data-level imaging property of
prostate MRI rather than a model-specific artifact.

\subsection{4.5 External validation on
Prostate158}\label{external-validation-on-prostate158-1}

On Prostate158, both models saturated near full sensitivity at the
PI-CAI-calibrated threshold (Table 5): bare 1.000 ± 0.000, refined 0.998
± 0.004; case-level specificity collapsed to near zero (bare 0.000 ±
0.000; refined 0.007 ± 0.016), consistent with an
apparent-diffusion-coefficient domain shift {[}4{]}; aggregate
discrimination was preserved (AUROC 0.655 ± 0.035 vs.~0.658 ± 0.025). At
matched sensitivity, descriptive specificity was 0.061 ± 0.053 (bare)
versus 0.050 ± 0.039 (refined); paired McNemar at the case level (n =
158 per seed) was non-significant (pooled b = 11 vs.~c = 14,
continuity-corrected \emph{p} = 0.69), with per-case bootstrap on
ΔCaseSpec −0.011 (95\% CI −0.039 to +0.018; \emph{p} = 0.52). A
sub-saturation sweep (targets 0.80--0.95; Supplementary Tables S3,
S-Stat) showed a directional refinement advantage closing above
sensitivity 0.90.

The P2b false-positive analysis replicated cross-center. Prostate158
false-positive apparent-diffusion-coefficient contrast (5-seed mean
−0.17 ± 0.06) matched true cancers (−0.17) and was separated from benign
tissue by an order of magnitude (+0.11). T2-weighted showed the same
ordering by median (false-positive −2.23 ± 0.56; cancer −0.97; benign
+0.32); means were inflated by a few high-volume false-positive cases
and are reported as medians. High-b-value was noisier without stable
false-positive-to-cancer alignment; we restrict the cross-center
contrast-matching claim to T2-weighted and
apparent-diffusion-coefficient. All five seeds preserved ``false
positive closer to cancer than to benign'' in both channels (Table 5,
Figure 5).

\section{Discussion}\label{discussion}

We report two findings. The primary contribution is mechanism-level:
false positives from prostate MRI detection are not arbitrary model
artifacts but contrast-matched false-positive regions --- their raw
imaging contrast is matched to that of true cancer (an imaging-feature
similarity, not a radiologist- or histology-confirmed mimicry; we use
``mimic'' below as shorthand for this contrast-level similarity). The
lesion-versus-benign and false-positive-versus-benign evidence
directions reproduced in 35 of 35 observations across five distinct
backbone architectures, in 15 of 15 across PI-CAI 5-fold × 3 seeds, in
105 of 105 across seven modality-perturbation scenarios, and in raw
T2-weighted and apparent-diffusion-coefficient contrast across an
external cohort (Cohen's d 1.10; FP/benign evidence ratio 2.38×) --- an
architecture-invariant, fold-invariant, cohort-spanning reproduction
indicating a data-level imaging property of prostate MRI. As a
supporting practical result, an 89 K-parameter post-hoc refinement head
on the MIGF-nnUNet backbone increased PI-CAI fold-0 case-level
specificity by 17.2\% relative (0.469 to 0.549; +0.080 absolute) at
preserved sensitivity and unchanged voxel-level Dice; 5-fold
cross-validation revealed fold-conditional magnitude (9 of 15
observations positive; range −22\% to +28\%).

Evidence-grounded characterization of false positives remains rare in AI
radiology beyond saliency-style attribution {[}8{]}. The FP-mimic
finding --- that false positives from a strong prostate MRI detector are
not random but resemble true cancers in raw imaging contrast far more
than they resemble benign tissue, with the direction reproducing 35/35
across five distinct architectures and 105/105 across
modality-perturbation scenarios --- reframes the specificity problem in
radiology AI from ``the model is wrong'' to ``the imaging signal is
genuinely ambiguous.'' Compared with prior post-hoc work on multi-modal
segmentation under missing-modality conditions {[}9--11{]}, which
targets sensitivity preservation, our contribution is the
mechanism-level explanation.

Limitations are as follows. First, although the case-level specificity
improvement of +17.2\% on fold-0 is consistent with subfield reporting
practice on PI-CAI, 5-fold cross-validation revealed fold-conditional
behavior (9 of 15 observations positive; range −22\% to +28\%); the
post-hoc residual is therefore sensitive to training-set composition,
and clinical deployment should validate the refinement effect on the
target cohort before integration. Second, the external cohort
(Prostate158, \emph{n} = 158) was acquired at a single non-PI-CAI site;
broader multi-site validation would strengthen generalizability. Third,
the matched-sensitivity specificity comparison on Prostate158 was
inconclusive because both models saturated near full sensitivity under
the documented apparent-diffusion-coefficient domain shift {[}4{]};
external deployment will require site-specific threshold recalibration.
Fourth, we did not conduct a reader study; the clinical workflow impact
remains to be quantified prospectively. Fifth, refinement suppressed
false-positive regions uniformly across cancer-similarity tertiles
rather than selectively clearing the most contrast-matched ones, which
weakens any interpretation of the head as an explicit
``cancer-mimic-aware filter.'' Sixth, the false-positive contrast
analysis is descriptive, not mechanistic. Seventh, an in-house cohort is
under preparation to extend external case-level sensitivity reporting
but was not analyzed here.

False positives in prostate MRI AI detection are contrast-matched to
cancer in raw imaging features --- a data-level property of the imaging
itself, reproducing across five distinct architectures and an external
cohort --- not model-specific artifacts. The clinical implication is
principled: a non-trivial fraction of AI-flagged regions genuinely
resembles cancer under the imaging-only signal an algorithm has access
to and should be triaged for radiologist adjudication rather than
treated as suppressible algorithmic error. Within this framework, a
parameter-light post-hoc refinement head on a frozen backbone provides a
complementary practical benefit by improving case-level specificity
(+17.2\% relative, +0.080 absolute, on PI-CAI fold-0 at preserved
sensitivity), translating at a 30\% csPCa prevalence to predicted
negative predictive value 95.0\% → 95.7\% and positive predictive value
43.2\% → 47.3\% (at 40\% prevalence, 92.5\% → 93.5\% and 54.2\% →
58.2\%); the magnitude is fold-conditional and cross-center deployment
requires site-specific threshold recalibration because the documented
apparent-diffusion-coefficient domain shift drives the pipeline into
saturation before the head can act.

\section{References}\label{references}

\protect\phantomsection\label{refs}

\clearpage

\section{Figures}\label{figures}

\includegraphics[width=0.85\linewidth,height=\textheight,keepaspectratio]{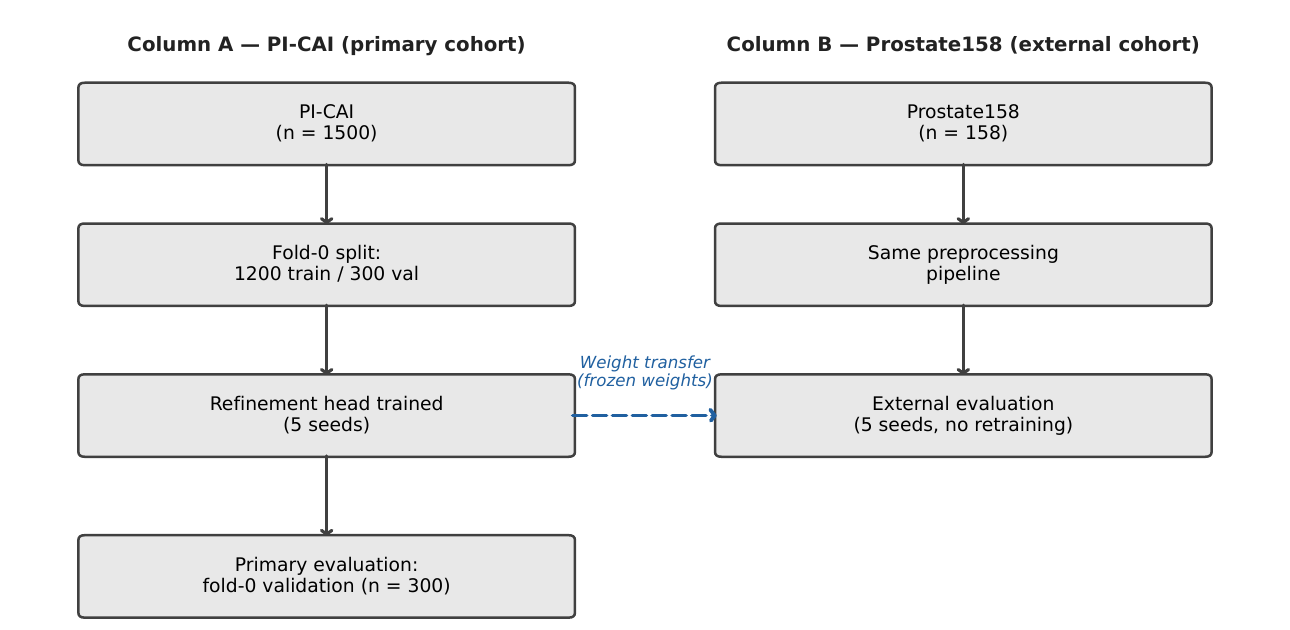}

\textbf{Figure 1.} Study flowchart. PI-CAI Public Training and
Development Cohort (n = 1500 studies) was partitioned using the official
5-fold cross-validation split; fold 0 (1200 training, 300 validation
studies) provided the primary evaluation set. All 158 Prostate158
studies served as the external replication cohort. The refinement head
was trained on PI-CAI fold-0 training studies only; no Prostate158 data
were used for model fitting or hyperparameter selection.

\clearpage

\includegraphics[width=0.85\linewidth,height=\textheight,keepaspectratio]{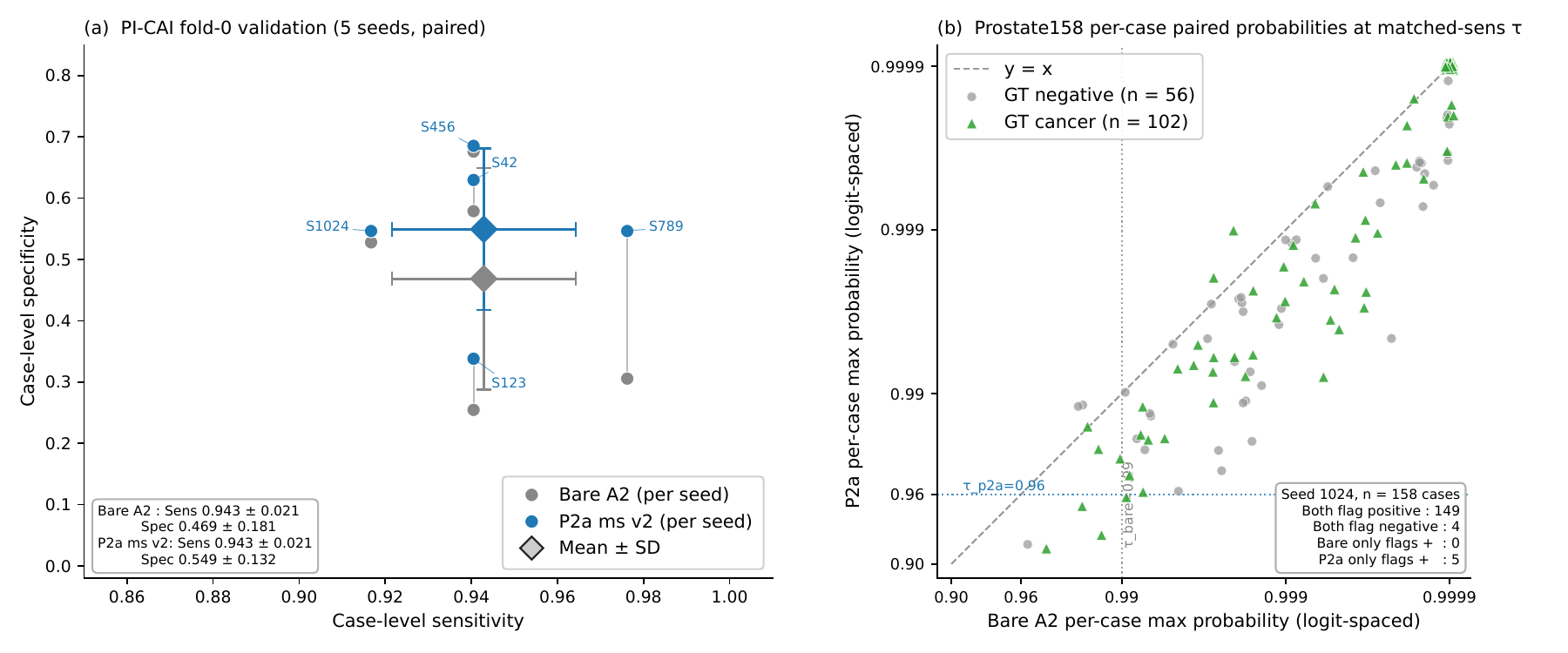}

\textbf{Figure 2.} Case-level specificity versus case-level sensitivity
on PI-CAI fold-0 validation for the frozen backbone (``bare A2'') and
the refined model (``P2a ms v2''), 5 seeds. Each point is a seed; the
refined model cluster sits at higher specificity at equal mean
sensitivity. Points are annotated with seed identifier for transparency.

\clearpage

\includegraphics[width=0.85\linewidth,height=\textheight,keepaspectratio]{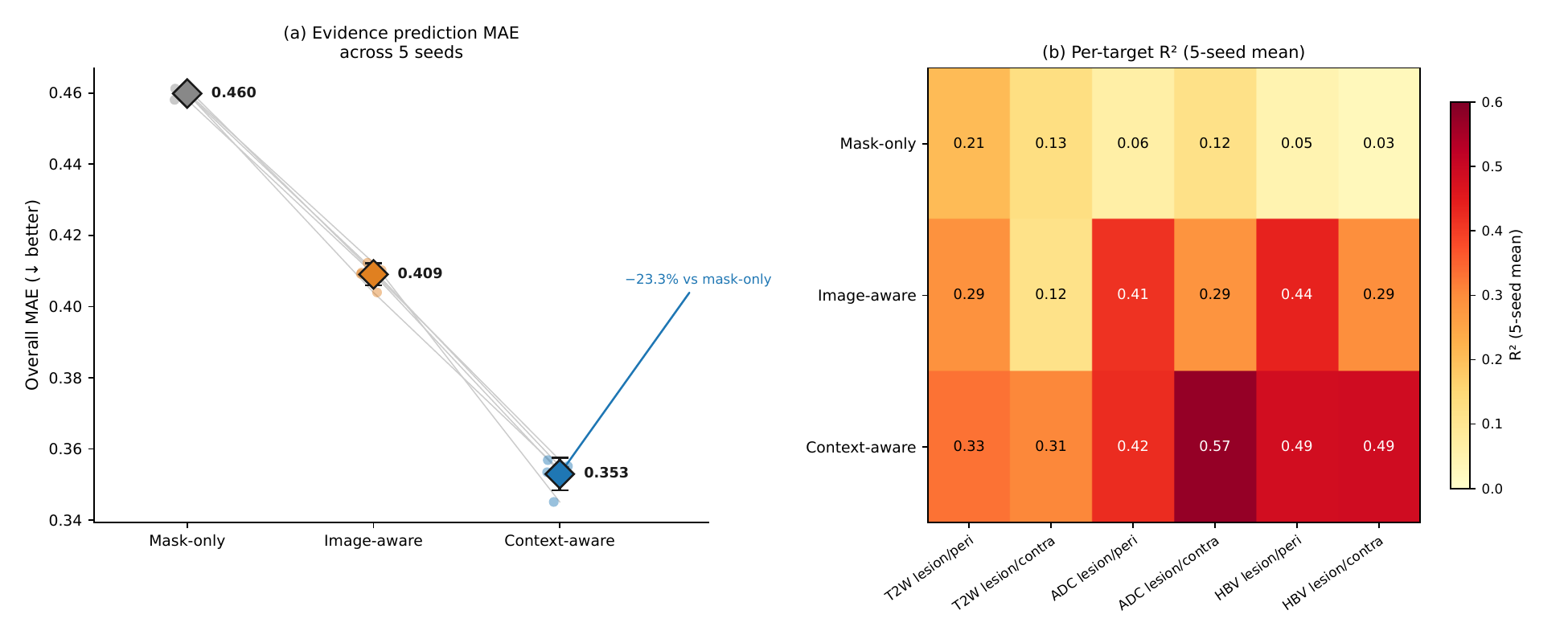}

\textbf{Figure 3.} Evidence prediction on PI-CAI fold-0 validation. (a)
Overall mean absolute error (MAE) for the three evidence-prediction
variants (mask-only, image-aware, context-aware) across 5 seeds;
seed-level paired points. (b) Per-target coefficient of determination
(\emph{R}\textsuperscript{2}) heatmap for the six continuous
contrast-ratio targets; darker indicates higher
\emph{R}\textsuperscript{2}.

\clearpage

\includegraphics[width=0.85\linewidth,height=\textheight,keepaspectratio]{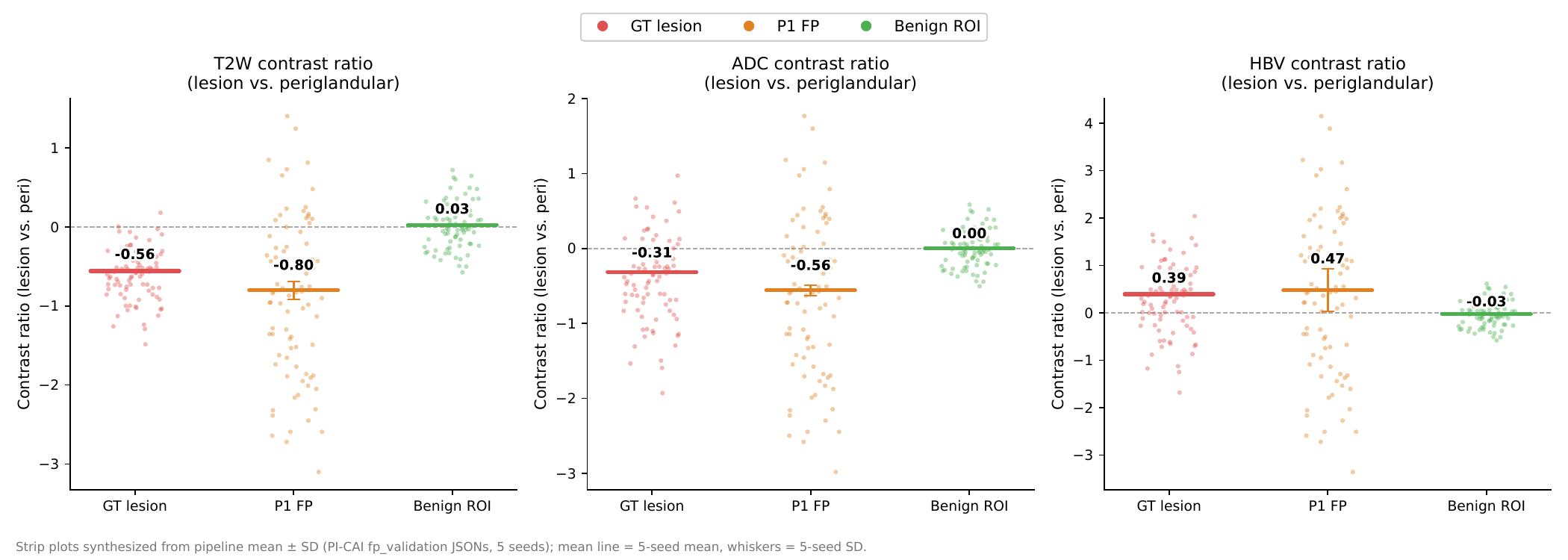}

\textbf{Figure 4.} Raw false-positive evidence analysis on PI-CAI.
Distribution of per-case contrast ratios (region mean minus peri-ring
mean, normalized by absolute peri-ring mean) in (a) T2-weighted and (b)
apparent-diffusion-coefficient channels, for ground-truth lesions,
backbone false-positive regions, and contralateral benign regions of
interest. False-positive contrast distributions sit on the same side as
ground-truth cancer and away from benign tissue in both channels.

\clearpage

\includegraphics[width=0.85\linewidth,height=\textheight,keepaspectratio]{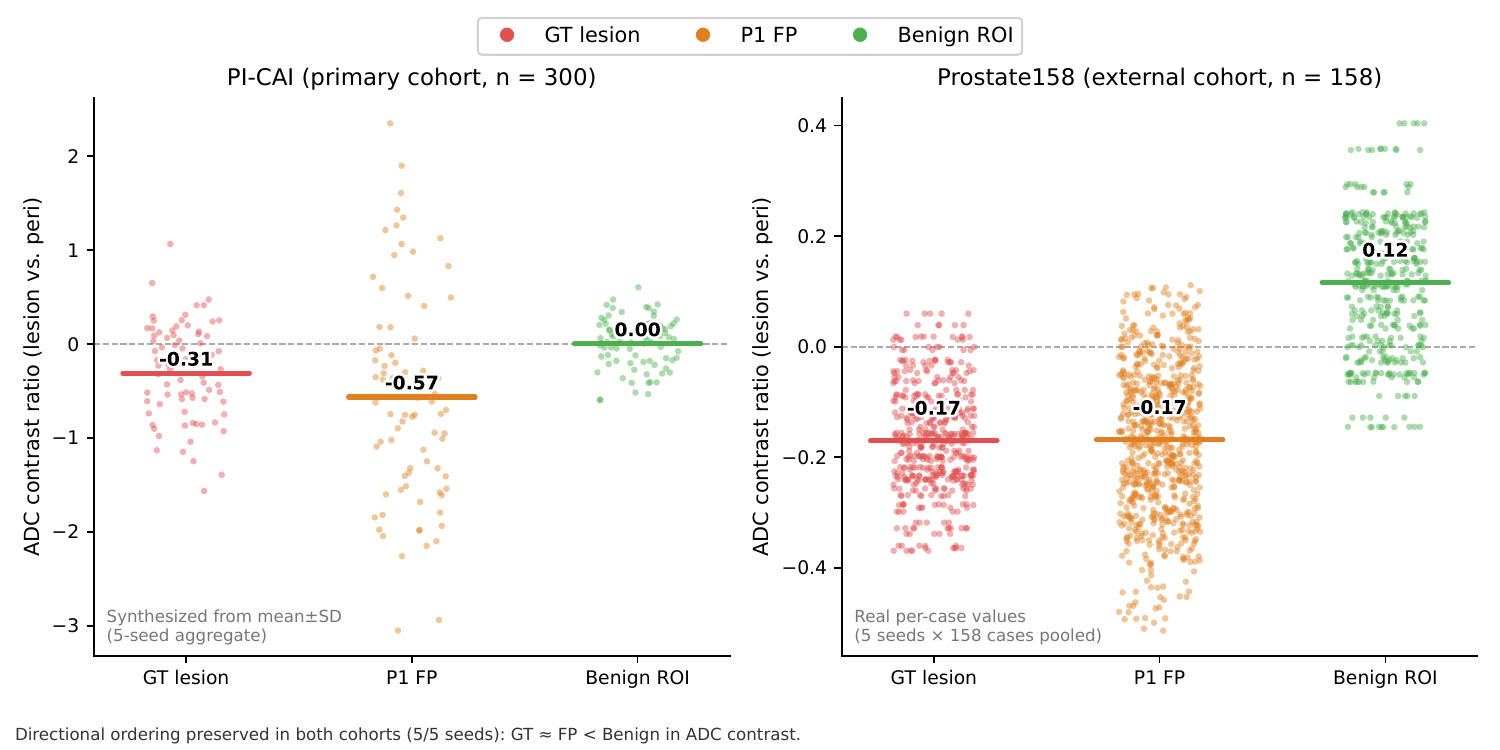}

\textbf{Figure 5.} Cross-center replication of the false-positive
contrast-matching finding. Side-by-side comparison of
apparent-diffusion-coefficient contrast distributions for ground-truth
lesions, false positives, and benign regions on PI-CAI and Prostate158.
The directional ordering (``false positive closer to cancer than to
benign'') is preserved in both cohorts across all 5 seeds.

\clearpage

\includegraphics[width=0.85\linewidth,height=\textheight,keepaspectratio]{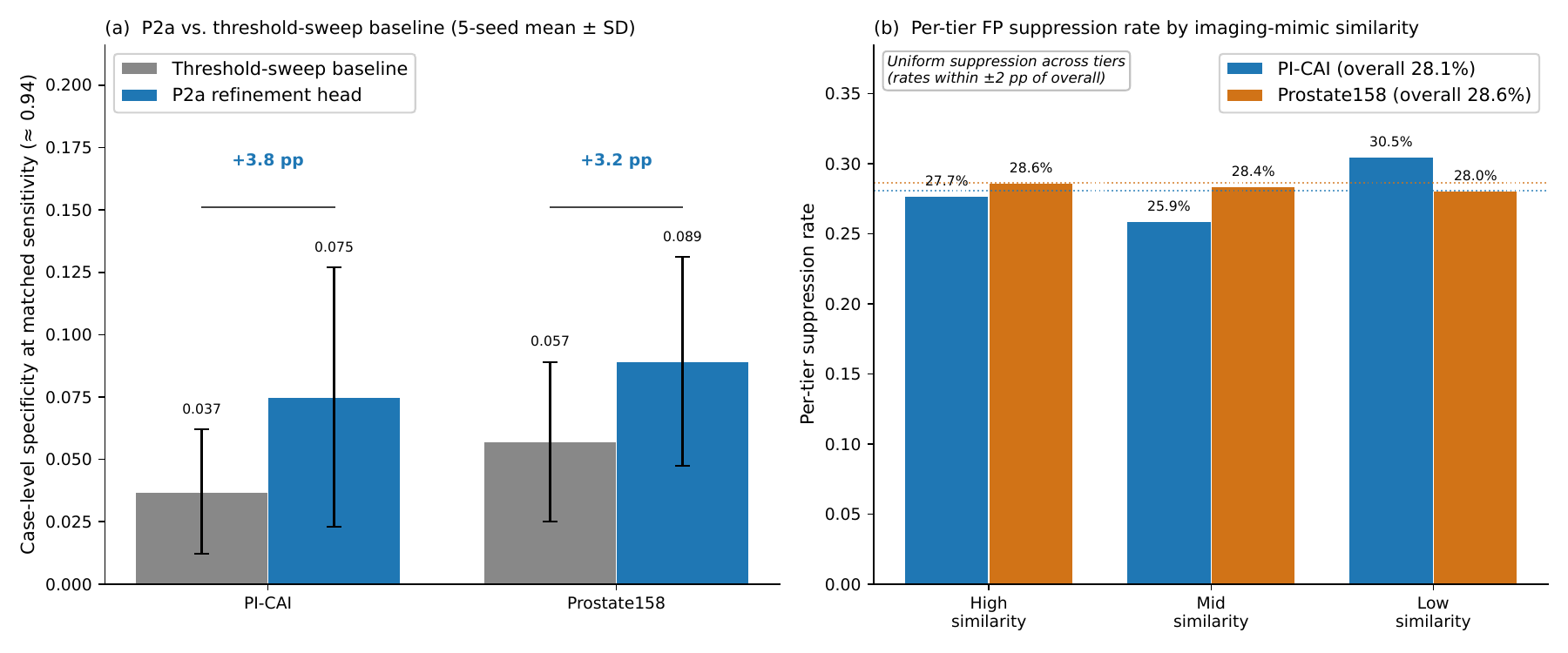}

\textbf{Figure 6.} (a) Matched-sensitivity case-level specificity on
PI-CAI and Prostate158 for the threshold-sweep baseline versus the P2a
refinement head (5-seed mean +/- sample standard deviation); the
+3.8-percentage-point advantage on PI-CAI is annotated. (b) Per-tier
false-positive suppression rate by cosine-similarity tertile (high, mid,
low) of the backbone's dec2 feature vector to the true-cancer centroid,
on both cohorts; per-tier rates lie within +/-2 percentage points of
each cohort's overall suppression rate, indicating uniform suppression
across the cancer-similarity spectrum.

\clearpage

\section{Tables}\label{tables}

\textbf{Table 1. Cohort demographics.}

{\def\LTcaptype{none} 
\begin{longtable}[]{@{}lll@{}}
\toprule\noalign{}
Characteristic & PI-CAI fold-0 val (n = 300) & Prostate158 (n = 158) \\
\midrule\noalign{}
\endhead
\bottomrule\noalign{}
\endlastfoot
Biopsy-confirmed csPCa, n (\%) & 84 (28\%) & 102 (65\%) \\
Pathology-negative, n (\%) & 216 (72\%) & 56 (35\%) \\
Median age, years (IQR) & 66 (62--71) & 64 (58--71) \\
Sex & All male & All male \\
Sequences provided & T2W, high-b DWI, ADC & T2W, high-b DWI, ADC \\
Source institution(s) & 3 Dutch hospitals & Single non-PI-CAI site \\
\end{longtable}
}

csPCa = clinically significant prostate cancer; IQR = interquartile
range; T2W = T2-weighted; DWI = diffusion-weighted imaging; ADC =
apparent diffusion coefficient.

\clearpage

\textbf{Table 2. P2a main comparison (PI-CAI fold-0 validation, 5
seeds).}

{\def\LTcaptype{none} 
\begin{longtable}[]{@{}llll@{}}
\toprule\noalign{}
Metric & Bare A2 & Refined (P2a ms v2) & Δ (refined − bare) \\
\midrule\noalign{}
\endhead
\bottomrule\noalign{}
\endlastfoot
PI-CAI Score & 0.730 ± 0.056 & 0.741 ± 0.047 & +0.011 \\
AUROC & 0.864 ± 0.047 & 0.878 ± 0.041 & +0.014 \\
Case-level sensitivity & 0.943 ± 0.021 & 0.943 ± 0.021 & 0.000 \\
Case-level specificity & 0.469 ± 0.181 & 0.549 ± 0.132 & +0.080 \\
Positive Dice & 0.487 ± 0.009 & 0.490 ± 0.005 & +0.003 \\
\end{longtable}
}

Values: 5-seed mean ± sample SD (ddof = 1); ideal-scenario operating
threshold. AUROC = area under the receiver operating characteristic
curve.

\clearpage

\textbf{Table 3. P2b evidence prediction performance (PI-CAI, 5 seeds).}

{\def\LTcaptype{none} 
\begin{longtable}[]{@{}
  >{\raggedright\arraybackslash}p{(\linewidth - 6\tabcolsep) * \real{0.2500}}
  >{\raggedright\arraybackslash}p{(\linewidth - 6\tabcolsep) * \real{0.2500}}
  >{\raggedright\arraybackslash}p{(\linewidth - 6\tabcolsep) * \real{0.2500}}
  >{\raggedright\arraybackslash}p{(\linewidth - 6\tabcolsep) * \real{0.2500}}@{}}
\toprule\noalign{}
\begin{minipage}[b]{\linewidth}\raggedright
Variant
\end{minipage} & \begin{minipage}[b]{\linewidth}\raggedright
Overall MAE (↓)
\end{minipage} & \begin{minipage}[b]{\linewidth}\raggedright
Suspicion macro-F1 (↑)
\end{minipage} & \begin{minipage}[b]{\linewidth}\raggedright
Δ MAE vs mask-only
\end{minipage} \\
\midrule\noalign{}
\endhead
\bottomrule\noalign{}
\endlastfoot
mask-only & 0.460 ± 0.001 & 0.617 ± 0.017 & --- \\
image-aware & 0.409 ± 0.003 & 0.608 ± 0.023 & −0.051 (−11.1\%) \\
context-aware & 0.353 ± 0.005 & 0.616 ± 0.034 & −0.107 (−23.3\%) \\
\end{longtable}
}

Values: 5-seed mean ± sample SD (ddof = 1). Ranking context-aware
\textless{} image-aware \textless{} mask-only preserved across all 5
seeds pairwise. MAE = mean absolute error; F1 = harmonic mean of
precision and recall.

\clearpage

\textbf{Table 4. False-positive contrast ratios on PI-CAI fold-0 (5-seed
aggregate).}

{\def\LTcaptype{none} 
\begin{longtable}[]{@{}
  >{\raggedright\arraybackslash}p{(\linewidth - 6\tabcolsep) * \real{0.2500}}
  >{\raggedright\arraybackslash}p{(\linewidth - 6\tabcolsep) * \real{0.2500}}
  >{\raggedright\arraybackslash}p{(\linewidth - 6\tabcolsep) * \real{0.2500}}
  >{\raggedright\arraybackslash}p{(\linewidth - 6\tabcolsep) * \real{0.2500}}@{}}
\toprule\noalign{}
\begin{minipage}[b]{\linewidth}\raggedright
MRI channel
\end{minipage} & \begin{minipage}[b]{\linewidth}\raggedright
GT lesion
\end{minipage} & \begin{minipage}[b]{\linewidth}\raggedright
P1 FP
\end{minipage} & \begin{minipage}[b]{\linewidth}\raggedright
Benign ROI
\end{minipage} \\
\midrule\noalign{}
\endhead
\bottomrule\noalign{}
\endlastfoot
T2W (lesion vs.~peri) & −0.556 ± 0.000 & −0.802 ± 0.113 & +0.027 ±
0.000 \\
ADC (lesion vs.~peri) & −0.312 ± 0.000 & −0.558 ± 0.071 & 0.000 ±
0.000 \\
HBV (lesion vs.~peri) & +0.393 ± 0.000 & +0.474 ± 0.448 & −0.027 ±
0.000 \\
\end{longtable}
}

Values: 5-seed mean ± SD (ddof=1 across 5 seed-level means). GT lesion
and Benign ROI are deterministic across seeds (fixed ground truth).
Contrast ratio = (ROI signal − peri-lesional ring signal) /
(\textbar peri-lesional ring signal\textbar{} + ε).

\textsuperscript{a} Paired Wilcoxon signed-rank test (P1 FP vs.~Benign
ROI, 5-seed mean as paired observations): T2W \emph{p} = 0.063; ADC
\emph{p} = 0.063; HBV \emph{p} = 0.063. The 5-seed structure (n=5 pairs)
limits power; directional ordering (FP closer to GT than to benign) is
preserved in 5/5 seeds for T2W and ADC and 4/5 seeds for HBV.

\textsuperscript{b} HBV FP values show high cross-seed variance (SD =
0.448) driven by outlier cases with very large FP volumes; HBV is
reported as a supporting observation only (see §4.4).

\textbf{Data source:} PI-CAI fp\_validation JSONs (5 seeds:
42/123/456/789/1024), \texttt{check1\_raw\_ratios} → per-seed group
means aggregated in Session 4. Per-case arrays not stored in JSON;
group-level aggregate mean ± within-group SD available per seed.

\clearpage

\textbf{Table 5. External validation on Prostate158 (5 seeds).}

{\def\LTcaptype{none} 
\begin{longtable}[]{@{}
  >{\raggedright\arraybackslash}p{(\linewidth - 4\tabcolsep) * \real{0.4747}}
  >{\raggedright\arraybackslash}p{(\linewidth - 4\tabcolsep) * \real{0.1919}}
  >{\raggedright\arraybackslash}p{(\linewidth - 4\tabcolsep) * \real{0.3333}}@{}}
\toprule\noalign{}
\begin{minipage}[b]{\linewidth}\raggedright
Metric / channel
\end{minipage} & \begin{minipage}[b]{\linewidth}\raggedright
Bare A2
\end{minipage} & \begin{minipage}[b]{\linewidth}\raggedright
Refined (P2a ms v2)
\end{minipage} \\
\midrule\noalign{}
\endhead
\bottomrule\noalign{}
\endlastfoot
\textbf{At PI-CAI-calibrated threshold} & & \\
PI-CAI Score & 0.417 ± 0.032 & 0.421 ± 0.042 \\
AUROC & 0.658 ± 0.025 & 0.655 ± 0.035 \\
Case-level sensitivity & 1.000 ± 0.000 & 0.998 ± 0.004 \\
Case-level specificity & 0.000 ± 0.000 & 0.007 ± 0.016 \\
\textbf{At matched sensitivity (target = PI-CAI CaseSens per seed)} &
& \\
Case-level specificity & 0.061 ± 0.053 & 0.050 ± 0.039 \\
Δ specificity (paired Wilcoxon \emph{p}) & & −0.011 ± 0.048 (\emph{p} =
0.62) \\
\textbf{FP contrast ratios (5-seed cross-case mean ± SD)} & & \\
T2W median: GT / FP / Benign & & −0.97 / −2.23 ± 0.56 / +0.32 \\
ADC mean: GT / FP / Benign & & −0.17 / −0.17 ± 0.06 / +0.11 \\
HBV (supporting only) & & noisy across seeds \\
\end{longtable}
}

Values: 5-seed mean ± sample SD (ddof = 1). AUROC = area under the
receiver operating characteristic curve.

\textbf{Footnote:} Values for bare and refined models in the FP contrast
section are equivalent because FP regions are defined from the frozen
backbone outputs. Directional ordering (GT ≈ FP \textless{} Benign for
ADC; GT \textless{} FP \textless{} Benign for T2W) was preserved in all
5 seeds, replicating the PI-CAI finding.

\clearpage

\section{Supplementary Tables}\label{supplementary-tables}

\textbf{Table S-5fold-Summary.} Five-fold cross-validation summary for
P2a (case-level specificity refinement) and P2b (evidence-grounded
false-positive analysis) across seven PI-CAI scenarios and the
Prostate158 external cohort. Each PI-CAI row aggregates 15 observations
(5 folds x 3 seeds: 42, 123, 789); the Prostate158 row aggregates 15
observations on the held-out cohort.

{\def\LTcaptype{none} 
\begin{longtable}[]{@{}
  >{\raggedright\arraybackslash}p{(\linewidth - 10\tabcolsep) * \real{0.0872}}
  >{\raggedleft\arraybackslash}p{(\linewidth - 10\tabcolsep) * \real{0.1795}}
  >{\raggedright\arraybackslash}p{(\linewidth - 10\tabcolsep) * \real{0.1949}}
  >{\raggedleft\arraybackslash}p{(\linewidth - 10\tabcolsep) * \real{0.1487}}
  >{\raggedleft\arraybackslash}p{(\linewidth - 10\tabcolsep) * \real{0.1744}}
  >{\raggedleft\arraybackslash}p{(\linewidth - 10\tabcolsep) * \real{0.2154}}@{}}
\toprule\noalign{}
\begin{minipage}[b]{\linewidth}\raggedright
Scenario
\end{minipage} & \begin{minipage}[b]{\linewidth}\raggedleft
P2a CaseSpec mean +/- SD (n = 15)
\end{minipage} & \begin{minipage}[b]{\linewidth}\raggedright
P2a Delta vs paper-reference
\end{minipage} & \begin{minipage}[b]{\linewidth}\raggedleft
P2b MAE mean +/- SD (n = 15)
\end{minipage} & \begin{minipage}[b]{\linewidth}\raggedleft
P2b lesion \textgreater{} benign reproduction
\end{minipage} & \begin{minipage}[b]{\linewidth}\raggedleft
P2b margin (lesion - benign) mean +/- SD
\end{minipage} \\
\midrule\noalign{}
\endhead
\bottomrule\noalign{}
\endlastfoot
PI-CAI ideal & 0.4926 +/- 0.150 & -0.057 vs fold-0 5-seed (0.549) &
0.344 +/- 0.014 & 15 / 15 & 0.297 +/- 0.035 \\
missing\_t2w & 0.1363 +/- 0.105 & (no fold-0 5-seed reference) & 0.426
+/- 0.017 & 15 / 15 & 0.247 +/- 0.057 \\
missing\_hbv & 0.4758 +/- 0.152 & -- & 0.405 +/- 0.015 & 15 / 15 & 0.155
+/- 0.023 \\
missing\_adc & 0.4808 +/- 0.131 & -- & 0.381 +/- 0.013 & 15 / 15 & 0.268
+/- 0.027 \\
artifact\_t2w & 0.6627 +/- 0.185 & -- & 0.347 +/- 0.015 & 15 / 15 &
0.278 +/- 0.031 \\
artifact\_hbv & 0.4923 +/- 0.156 & -- & 0.347 +/- 0.013 & 15 / 15 &
0.287 +/- 0.037 \\
artifact\_adc & 0.4486 +/- 0.144 & -- & 0.348 +/- 0.012 & 15 / 15 &
0.287 +/- 0.032 \\
Prostate158 & 0.0036 +/- 0.010 & (architecture-conditional collapse) &
0.373 +/- 0.029 & (cross-cohort; see Table 5) & (cross-cohort; see Table
5) \\
\end{longtable}
}

\textbf{Footnote.} P2a Delta vs paper-reference reports the difference
relative to the Paper 2 fold-0 5-seed reference (0.549) on PI-CAI ideal;
the remaining six PI-CAI scenarios do not have a 5-seed reference in
this manuscript. P2b ``lesion \textgreater{} benign reproduction''
tallies the number of (fold, seed) observations in which the
evidence-head lesion mean exceeded the benign mean. Across the seven
PI-CAI scenarios x 15 observations, lesion \textgreater{} benign
reproduces in 105 / 105. Source: the P2a and P2b 5-fold audit-dump
summaries.

\clearpage

\textbf{Table S-5fold-PerFold.} Per-(fold, seed) P2a case-level
specificity on PI-CAI ideal scenario. Comparison against the Paper 1 A2
backbone 5-fold case-level specificity baseline (0.4562). Bold per-fold
mean = match-or-exceed the Paper 2 fold-0 5-seed reference (+17.2\% on
PI-CAI fold-0, CaseSpec 0.549; equivalently, fold mean ≥ 0.5347).

{\def\LTcaptype{none} 
\begin{longtable}[]{@{}
  >{\raggedleft\arraybackslash}p{(\linewidth - 12\tabcolsep) * \real{0.0723}}
  >{\raggedleft\arraybackslash}p{(\linewidth - 12\tabcolsep) * \real{0.1084}}
  >{\raggedleft\arraybackslash}p{(\linewidth - 12\tabcolsep) * \real{0.1205}}
  >{\raggedleft\arraybackslash}p{(\linewidth - 12\tabcolsep) * \real{0.1205}}
  >{\raggedleft\arraybackslash}p{(\linewidth - 12\tabcolsep) * \real{0.1325}}
  >{\raggedleft\arraybackslash}p{(\linewidth - 12\tabcolsep) * \real{0.3253}}
  >{\raggedleft\arraybackslash}p{(\linewidth - 12\tabcolsep) * \real{0.1205}}@{}}
\toprule\noalign{}
\begin{minipage}[b]{\linewidth}\raggedleft
Fold
\end{minipage} & \begin{minipage}[b]{\linewidth}\raggedleft
Seed 42
\end{minipage} & \begin{minipage}[b]{\linewidth}\raggedleft
Seed 123
\end{minipage} & \begin{minipage}[b]{\linewidth}\raggedleft
Seed 789
\end{minipage} & \begin{minipage}[b]{\linewidth}\raggedleft
Fold mean
\end{minipage} & \begin{minipage}[b]{\linewidth}\raggedleft
Δ vs A2 baseline (0.4562)
\end{minipage} & \begin{minipage}[b]{\linewidth}\raggedleft
Relative
\end{minipage} \\
\midrule\noalign{}
\endhead
\bottomrule\noalign{}
\endlastfoot
0 & 0.6296 & 0.4444 & 0.5648 & \textbf{0.5463} & +0.0901 &
\textbf{+19.7\%} \\
1 & 0.5619 & 0.5841 & 0.5531 & \textbf{0.5664} & +0.1102 &
\textbf{+24.1\%} \\
2 & 0.3272 & 0.5484 & 0.3641 & 0.4132 & −0.0430 & −9.4\% \\
3 & 0.3921 & 0.6079 & 0.7489 & \textbf{0.5830} & +0.1268 &
\textbf{+27.8\%} \\
4 & 0.3810 & 0.5381 & 0.1429 & 0.3540 & −0.1022 & −22.4\% \\
\end{longtable}
}

\textbf{Footnote.} Per-seed CaseSpec extracted from the 15 per-(fold,
seed) result files; computed fold means match the audit-dump summary to
four decimal places. Seeds used in the 5-fold supplementary were 42,
123, 789 (the original Paper 2 fold-0 reference used seeds 42, 123, 456,
789, 1024). The 0.4562 baseline is the Paper 1 A2 5-fold mean case-level
specificity. Direction sign across all 15 observations: 9 / 15 positive,
6 / 15 negative; range {[}−0.143 (fold 4 seed 789, relative −31.3\%) to
+0.293 (fold 3 seed 789, relative +64.2\%){]} versus the per-fold A2
baseline.

\clearpage

\textbf{Table S-35backbones.} Backbone-agnostic verification of the P2b
false-positive contrast-matching mechanism. The context-aware evidence
head (identical architecture and hyperparameters) was trained on five
backbones across all 5 PI-CAI folds. Observation counts: the four added
backbones (bare nnU-Net, bare U-Net, bare Mamba, MIGF-Mamba A2)
contribute 20 observations (4 backbones x 5 folds at seed 42);
MIGF-nnUNet A2 contributes its full 15-observation reference (5 folds x
3 seeds: 42, 123, 789); total = 35. For readability the sub-tables below
first group the seed-42 rows of all five backbones (a 25-observation
block), then add the MIGF-nnUNet A2 seed 123 and 789 supplement (10
observations), summing to 35. Both directions reproduce in 35 / 35
observations across five distinct architectures.

\textbf{Sub-table A. Lesion-evidence \textgreater{} benign-evidence
reproduction.}

{\def\LTcaptype{none} 
\begin{longtable}[]{@{}
  >{\raggedright\arraybackslash}p{(\linewidth - 12\tabcolsep) * \real{0.2892}}
  >{\centering\arraybackslash}p{(\linewidth - 12\tabcolsep) * \real{0.0964}}
  >{\centering\arraybackslash}p{(\linewidth - 12\tabcolsep) * \real{0.0964}}
  >{\centering\arraybackslash}p{(\linewidth - 12\tabcolsep) * \real{0.0964}}
  >{\centering\arraybackslash}p{(\linewidth - 12\tabcolsep) * \real{0.0964}}
  >{\centering\arraybackslash}p{(\linewidth - 12\tabcolsep) * \real{0.0964}}
  >{\raggedleft\arraybackslash}p{(\linewidth - 12\tabcolsep) * \real{0.2289}}@{}}
\toprule\noalign{}
\begin{minipage}[b]{\linewidth}\raggedright
Backbone
\end{minipage} & \begin{minipage}[b]{\linewidth}\centering
Fold 0
\end{minipage} & \begin{minipage}[b]{\linewidth}\centering
Fold 1
\end{minipage} & \begin{minipage}[b]{\linewidth}\centering
Fold 2
\end{minipage} & \begin{minipage}[b]{\linewidth}\centering
Fold 3
\end{minipage} & \begin{minipage}[b]{\linewidth}\centering
Fold 4
\end{minipage} & \begin{minipage}[b]{\linewidth}\raggedleft
Reproductions / 5
\end{minipage} \\
\midrule\noalign{}
\endhead
\bottomrule\noalign{}
\endlastfoot
bare nnU-Net & 1 & 1 & 1 & 1 & 1 & 5/5 \\
bare U-Net & 1 & 1 & 1 & 1 & 1 & 5/5 \\
bare Mamba & 1 & 1 & 1 & 1 & 1 & 5/5 \\
MIGF-Mamba A2 & 1 & 1 & 1 & 1 & 1 & 5/5 \\
MIGF-nnUNet A2 (seed 42 subset) & 1 & 1 & 1 & 1 & 1 & 5/5 \\
\textbf{Sub-total} & & & & & & 25/25 \\
MIGF-nnUNet A2 (seeds 123, 789 supplement) & & & & & & 10/10 \\
\textbf{TOTAL} & & & & & & \textbf{35/35} \\
\end{longtable}
}

\textbf{Sub-table B. False-positive-evidence \textgreater{}
benign-evidence reproduction.}

{\def\LTcaptype{none} 
\begin{longtable}[]{@{}
  >{\raggedright\arraybackslash}p{(\linewidth - 12\tabcolsep) * \real{0.2892}}
  >{\centering\arraybackslash}p{(\linewidth - 12\tabcolsep) * \real{0.0964}}
  >{\centering\arraybackslash}p{(\linewidth - 12\tabcolsep) * \real{0.0964}}
  >{\centering\arraybackslash}p{(\linewidth - 12\tabcolsep) * \real{0.0964}}
  >{\centering\arraybackslash}p{(\linewidth - 12\tabcolsep) * \real{0.0964}}
  >{\centering\arraybackslash}p{(\linewidth - 12\tabcolsep) * \real{0.0964}}
  >{\raggedleft\arraybackslash}p{(\linewidth - 12\tabcolsep) * \real{0.2289}}@{}}
\toprule\noalign{}
\begin{minipage}[b]{\linewidth}\raggedright
Backbone
\end{minipage} & \begin{minipage}[b]{\linewidth}\centering
Fold 0
\end{minipage} & \begin{minipage}[b]{\linewidth}\centering
Fold 1
\end{minipage} & \begin{minipage}[b]{\linewidth}\centering
Fold 2
\end{minipage} & \begin{minipage}[b]{\linewidth}\centering
Fold 3
\end{minipage} & \begin{minipage}[b]{\linewidth}\centering
Fold 4
\end{minipage} & \begin{minipage}[b]{\linewidth}\raggedleft
Reproductions / 5
\end{minipage} \\
\midrule\noalign{}
\endhead
\bottomrule\noalign{}
\endlastfoot
bare nnU-Net & 1 & 1 & 1 & 1 & 1 & 5/5 \\
bare U-Net & 1 & 1 & 1 & 1 & 1 & 5/5 \\
bare Mamba & 1 & 1 & 1 & 1 & 1 & 5/5 \\
MIGF-Mamba A2 & 1 & 1 & 1 & 1 & 1 & 5/5 \\
MIGF-nnUNet A2 (seed 42 subset) & 1 & 1 & 1 & 1 & 1 & 5/5 \\
\textbf{Sub-total} & & & & & & 25/25 \\
MIGF-nnUNet A2 (seeds 123, 789 supplement) & & & & & & 10/10 \\
\textbf{TOTAL} & & & & & & \textbf{35/35} \\
\end{longtable}
}

\textbf{Footnote.} The four added backbones (bare nnU-Net, bare U-Net,
bare Mamba, MIGF-Mamba A2) contribute 20 observations (4 backbones x 5
folds x seed 42); MIGF-nnUNet A2 contributes its full 5-fold x 3-seed
P2b reference of 15 observations. Combined, 35 / 35 observations
replicate both directions (lesion \textgreater{} benign and
false-positive \textgreater{} benign). Cohen's d on lesion-vs-benign
margin pooled across MIGF-nnUNet A2 5-fold: 1.10 (range 0.91-1.36). FP /
benign evidence ratio mean: 2.38x (range 1.91x-2.69x). The P2b
context-aware evidence head, hyperparameters, and 100-epoch Adam (lr =
1e-4) training protocol are identical across all five backbones.
Implication: the false-positive contrast-matching phenomenon (false
positives sharing raw imaging contrast with true cancer) is a data-level
imaging property of prostate MRI, not a backbone-specific artifact.

\clearpage

\textbf{Table S-Baselines.} Per-seed case-level specificity at matched
case-level sensitivity (target ≈ 0.94) for three post-hoc baseline
strategies versus the P2a refinement head, on PI-CAI fold-0 validation
(n = 300) and Prostate158 (n = 158).

{\def\LTcaptype{none} 
\begin{longtable}[]{@{}
  >{\raggedright\arraybackslash}p{(\linewidth - 24\tabcolsep) * \real{0.1429}}
  >{\raggedleft\arraybackslash}p{(\linewidth - 24\tabcolsep) * \real{0.0816}}
  >{\raggedleft\arraybackslash}p{(\linewidth - 24\tabcolsep) * \real{0.0476}}
  >{\raggedleft\arraybackslash}p{(\linewidth - 24\tabcolsep) * \real{0.0476}}
  >{\raggedleft\arraybackslash}p{(\linewidth - 24\tabcolsep) * \real{0.0476}}
  >{\raggedleft\arraybackslash}p{(\linewidth - 24\tabcolsep) * \real{0.0476}}
  >{\raggedleft\arraybackslash}p{(\linewidth - 24\tabcolsep) * \real{0.1224}}
  >{\raggedleft\arraybackslash}p{(\linewidth - 24\tabcolsep) * \real{0.1156}}
  >{\raggedleft\arraybackslash}p{(\linewidth - 24\tabcolsep) * \real{0.0476}}
  >{\raggedleft\arraybackslash}p{(\linewidth - 24\tabcolsep) * \real{0.0476}}
  >{\raggedleft\arraybackslash}p{(\linewidth - 24\tabcolsep) * \real{0.0476}}
  >{\raggedleft\arraybackslash}p{(\linewidth - 24\tabcolsep) * \real{0.0476}}
  >{\raggedleft\arraybackslash}p{(\linewidth - 24\tabcolsep) * \real{0.1565}}@{}}
\toprule\noalign{}
\begin{minipage}[b]{\linewidth}\raggedright
Method
\end{minipage} & \begin{minipage}[b]{\linewidth}\raggedleft
PI-CAI s42
\end{minipage} & \begin{minipage}[b]{\linewidth}\raggedleft
s123
\end{minipage} & \begin{minipage}[b]{\linewidth}\raggedleft
s456
\end{minipage} & \begin{minipage}[b]{\linewidth}\raggedleft
s789
\end{minipage} & \begin{minipage}[b]{\linewidth}\raggedleft
s1024
\end{minipage} & \begin{minipage}[b]{\linewidth}\raggedleft
PI-CAI Mean ± SD
\end{minipage} & \begin{minipage}[b]{\linewidth}\raggedleft
Prostate158 s42
\end{minipage} & \begin{minipage}[b]{\linewidth}\raggedleft
s123
\end{minipage} & \begin{minipage}[b]{\linewidth}\raggedleft
s456
\end{minipage} & \begin{minipage}[b]{\linewidth}\raggedleft
s789
\end{minipage} & \begin{minipage}[b]{\linewidth}\raggedleft
s1024
\end{minipage} & \begin{minipage}[b]{\linewidth}\raggedleft
Prostate158 Mean ± SD
\end{minipage} \\
\midrule\noalign{}
\endhead
\bottomrule\noalign{}
\endlastfoot
Temperature scaling & 0.042 & 0.000 & 0.000 & 0.000 & 0.000 & 0.008 ±
0.019 & 0.026 & --- & --- & --- & --- & 0.026 (n = 1) \\
Platt scaling & 0.042 & 0.069 & 0.146 & 0.139 & 0.097 & 0.099 ± 0.045 &
0.026 & --- & --- & --- & --- & 0.026 (n = 1) \\
Threshold sweep & 0.023 & 0.060 & 0.000 & 0.051 & 0.051 & 0.037 ± 0.025
& 0.089 & 0.036 & 0.018 & 0.089 & 0.054 & 0.057 ± 0.032 \\
P2a refinement & 0.042 & 0.106 & 0.000 & 0.106 & 0.120 & 0.075 ± 0.052 &
0.125 & 0.054 & 0.107 & 0.125 & 0.036 & 0.089 ± 0.042 \\
\end{longtable}
}

\textbf{Footnote.} At matched sensitivity (target 0.94 ± 0.02).
Temperature and Platt scaling are shown here for completeness; under our
case-level decision rule (connected component ≥ 10 voxels), monotonic
recalibration preserves the ROC and yields identical CaseSpec to
threshold sweep. Temperature and Platt scaling values on Prostate158 are
single-seed (seed 42) because the remaining four seeds degenerate to T →
∞ or produce ROC-equivalent operating points (see
experiments/baselines/DECISION.md). Per-seed deltas (P2a − threshold
sweep): PI-CAI +0.019 / +0.046 / 0.000 / +0.056 / +0.069 (mean +0.038 ±
0.028); Prostate158 +0.036 / +0.018 / 0.000 / −0.018 / +0.089 (mean
+0.032 ± 0.039). Seed 456 on PI-CAI is the saturated all-positive
threshold seed for which no method can reduce false positives without
violating matched sensitivity.

\clearpage

\textbf{Supplementary Table S3.} Case-level specificity on Prostate158
at four discrete sensitivity targets, for the bare backbone (A2) and the
refined model (P2a, multi-scale v2), 5 seeds. ``Reached'' means the
per-seed operating curve achieved case-level sensitivity within ±0.02 of
the target. Values in the ``mean CaseSpec (reached)'' columns are 5-seed
mean ± sample standard deviation (ddof = 1) restricted to seeds that
reached the target; ``n/5 reached'' reports the number of seeds.
Near-saturation targets (0.90, 0.95) produce sparse reached counts on
both models.

{\def\LTcaptype{none} 
\begin{longtable}[]{@{}
  >{\raggedright\arraybackslash}p{(\linewidth - 8\tabcolsep) * \real{0.2000}}
  >{\raggedright\arraybackslash}p{(\linewidth - 8\tabcolsep) * \real{0.2000}}
  >{\raggedright\arraybackslash}p{(\linewidth - 8\tabcolsep) * \real{0.2000}}
  >{\raggedright\arraybackslash}p{(\linewidth - 8\tabcolsep) * \real{0.2000}}
  >{\raggedright\arraybackslash}p{(\linewidth - 8\tabcolsep) * \real{0.2000}}@{}}
\toprule\noalign{}
\begin{minipage}[b]{\linewidth}\raggedright
Target CaseSens
\end{minipage} & \begin{minipage}[b]{\linewidth}\raggedright
Bare CaseSpec (reached)
\end{minipage} & \begin{minipage}[b]{\linewidth}\raggedright
Bare n/5 reached
\end{minipage} & \begin{minipage}[b]{\linewidth}\raggedright
Refined CaseSpec (reached)
\end{minipage} & \begin{minipage}[b]{\linewidth}\raggedright
Refined n/5 reached
\end{minipage} \\
\midrule\noalign{}
\endhead
\bottomrule\noalign{}
\endlastfoot
0.80 & 0.25 & 1/5 & 0.28 ± 0.01 & 2/5 \\
0.85 & 0.18 & 1/5 & 0.24 ± 0.04 & 2/5 \\
0.90 & 0.13 ± 0.03 & 2/5 & 0.11 ± 0.08 & 2/5 \\
0.95 & 0.06 ± 0.03 & 3/5 & 0.07 ± 0.03 & 4/5 \\
\end{longtable}
}

Source:
\texttt{experiments/prostate158\_sensspec\_curve/sensspec\_aggregate.json},
2026-04-20.

\clearpage

\textbf{Table S-Stat.} Seed-level paired Wilcoxon signed-rank analysis
of matched-sensitivity case-level specificity gain (P2a − bare A2),
retained as a sensitivity check to the main-text case-level inference
(§3.7).

{\def\LTcaptype{none} 
\begin{longtable}[]{@{}
  >{\raggedright\arraybackslash}p{(\linewidth - 8\tabcolsep) * \real{0.1413}}
  >{\raggedright\arraybackslash}p{(\linewidth - 8\tabcolsep) * \real{0.1957}}
  >{\raggedright\arraybackslash}p{(\linewidth - 8\tabcolsep) * \real{0.2935}}
  >{\raggedright\arraybackslash}p{(\linewidth - 8\tabcolsep) * \real{0.1196}}
  >{\raggedright\arraybackslash}p{(\linewidth - 8\tabcolsep) * \real{0.2500}}@{}}
\toprule\noalign{}
\begin{minipage}[b]{\linewidth}\raggedright
Cohort
\end{minipage} & \begin{minipage}[b]{\linewidth}\raggedright
Metric
\end{minipage} & \begin{minipage}[b]{\linewidth}\raggedright
5-seed per-seed values
\end{minipage} & \begin{minipage}[b]{\linewidth}\raggedright
Mean ± SD
\end{minipage} & \begin{minipage}[b]{\linewidth}\raggedright
Wilcoxon signed-rank one-sided p (H1: P2a \textgreater{} bare)
\end{minipage} \\
\midrule\noalign{}
\endhead
\bottomrule\noalign{}
\endlastfoot
PI-CAI fold-0 val & ΔCaseSpec at matched sens & +0.019, +0.046, 0.000,
+0.056, +0.069 & +0.038 ± 0.028 & p = 0.06 \\
Prostate158 & ΔCaseSpec at matched sens & +0.036, +0.018, 0.000, −0.018,
−0.089 & −0.011 ± 0.048 & p = 0.62 \\
\end{longtable}
}

\textbf{Footnote.} Seed-level Wilcoxon retained as sensitivity analysis;
main-text inference uses case-level McNemar + bootstrap per §3.7.
Per-seed ΔCaseSpec on PI-CAI from
\texttt{experiments/p2a\_5fold\_cv/bootstrap\_results.json}; per-seed
ΔCaseSpec on Prostate158 from
\texttt{experiments/prostate158\_matched\_sens/matched\_sens\_aggregate.json}.
Wilcoxon one-sided p was computed with
\texttt{scipy.stats.wilcoxon(alternative="greater",\ zero\_method="zsplit")}
(handles the tied zero at PI-CAI seed 456). The PI-CAI p = 0.06
approaches but does not cross α = 0.05 at n = 5 seeds, consistent with
the n = 5 Wilcoxon structural minimum p ≈ 0.03 and illustrating why the
main-text inference is case-level rather than seed-level.

\clearpage

\textbf{Table S-Bootstrap.} Per-seed case-level paired bootstrap
confidence intervals for the matched-sensitivity case-level specificity
gain (P2a − bare A2) on Prostate158 negative cases (n = 56 per seed).
Pooled row aggregates all 5 seeds × 56 negative cases = 280
case-observations.

{\def\LTcaptype{none} 
\begin{longtable}[]{@{}
  >{\raggedright\arraybackslash}p{(\linewidth - 12\tabcolsep) * \real{0.1986}}
  >{\raggedright\arraybackslash}p{(\linewidth - 12\tabcolsep) * \real{0.0922}}
  >{\raggedleft\arraybackslash}p{(\linewidth - 12\tabcolsep) * \real{0.1277}}
  >{\raggedleft\arraybackslash}p{(\linewidth - 12\tabcolsep) * \real{0.1986}}
  >{\raggedleft\arraybackslash}p{(\linewidth - 12\tabcolsep) * \real{0.0993}}
  >{\raggedleft\arraybackslash}p{(\linewidth - 12\tabcolsep) * \real{0.0993}}
  >{\raggedleft\arraybackslash}p{(\linewidth - 12\tabcolsep) * \real{0.1844}}@{}}
\toprule\noalign{}
\begin{minipage}[b]{\linewidth}\raggedright
Seed
\end{minipage} & \begin{minipage}[b]{\linewidth}\raggedright
Cohort
\end{minipage} & \begin{minipage}[b]{\linewidth}\raggedleft
n negative cases
\end{minipage} & \begin{minipage}[b]{\linewidth}\raggedleft
Δ CaseSpec (per-case mean)
\end{minipage} & \begin{minipage}[b]{\linewidth}\raggedleft
95\% CI lower
\end{minipage} & \begin{minipage}[b]{\linewidth}\raggedleft
95\% CI upper
\end{minipage} & \begin{minipage}[b]{\linewidth}\raggedleft
p (two-sided, bootstrap)
\end{minipage} \\
\midrule\noalign{}
\endhead
\bottomrule\noalign{}
\endlastfoot
42 & Prostate158 & 56 & +0.036 & 0.000 & +0.089 & 0.268 \\
123 & Prostate158 & 56 & +0.018 & 0.000 & +0.054 & 0.721 \\
456 & Prostate158 & 56 & 0.000 & 0.000 & 0.000 & 1.000 \\
789 & Prostate158 & 56 & −0.018 & −0.071 & +0.036 & 0.782 \\
1024 & Prostate158 & 56 & −0.089 & −0.196 & 0.000 & 0.112 \\
Pooled (5 seeds × 56) & Prostate158 & 280 & −0.011 & −0.039 & +0.018 &
0.517 \\
\end{longtable}
}

\textbf{Footnote.} Per-case paired bootstrap with 10,000 resamples (rng
seed 20260421). Seed-level estimates pooled across all 5 seeds × 56
negative cases = 280 case-observations; the pooled resampler draws with
replacement from the pooled (seed, case) observation index. The PI-CAI
per-case detection sidecars required to run the analogous analysis on
PI-CAI fold-0 validation were not available in the local experiments
tree; the PI-CAI bootstrap reported in the main text (§3.7) is therefore
the non-paired per-seed cohort resampling from
\texttt{experiments/p2a\_5fold\_cv/bootstrap\_results.json}. Seed 456 is
the saturated all-positive seed for which Δ = 0 by construction.

\clearpage

\textbf{Table S-Stratification.} Pooled-across-5-seeds false-positive
counts per similarity tier and per-tier suppression rates for PI-CAI
fold-0 validation and Prostate158. Tiers defined by tertiles of cosine
similarity between the FP-region contrast-ratio vector and the
case-specific ground-truth cancer contrast-ratio vector. Negative-case
FPs (FPs in patients with no ground-truth cancer) have no case-specific
cancer vector and are reported separately.

{\def\LTcaptype{none} 
\begin{longtable}[]{@{}
  >{\raggedright\arraybackslash}p{(\linewidth - 8\tabcolsep) * \real{0.1461}}
  >{\raggedright\arraybackslash}p{(\linewidth - 8\tabcolsep) * \real{0.2584}}
  >{\raggedleft\arraybackslash}p{(\linewidth - 8\tabcolsep) * \real{0.2584}}
  >{\raggedleft\arraybackslash}p{(\linewidth - 8\tabcolsep) * \real{0.1348}}
  >{\raggedleft\arraybackslash}p{(\linewidth - 8\tabcolsep) * \real{0.2022}}@{}}
\toprule\noalign{}
\begin{minipage}[b]{\linewidth}\raggedright
Cohort
\end{minipage} & \begin{minipage}[b]{\linewidth}\raggedright
Tier
\end{minipage} & \begin{minipage}[b]{\linewidth}\raggedleft
n FPs (5-seed pooled)
\end{minipage} & \begin{minipage}[b]{\linewidth}\raggedleft
\% of total
\end{minipage} & \begin{minipage}[b]{\linewidth}\raggedleft
Suppression rate
\end{minipage} \\
\midrule\noalign{}
\endhead
\bottomrule\noalign{}
\endlastfoot
PI-CAI & High similarity & 1,401 & 9.8\% & 27.7\% \\
PI-CAI & Mid similarity & 1,401 & 9.8\% & 25.9\% \\
PI-CAI & Low similarity & 1,401 & 9.8\% & 30.5\% \\
PI-CAI & Negative-case FPs & 10,071 & 70.5\% & 28.1\% \\
Prostate158 & High similarity & 1,494 & 22.0\% & 28.6\% \\
Prostate158 & Mid similarity & 1,494 & 22.0\% & 28.4\% \\
Prostate158 & Low similarity & 1,494 & 22.0\% & 28.0\% \\
Prostate158 & Negative-case FPs & 2,313 & 34.0\% & 29.1\% \\
\end{longtable}
}

\textbf{Footnote.} Tiers defined by tertiles of cosine similarity
between FP-region contrast-ratio vector and case-specific GT cancer
contrast-ratio vector. Uniform suppression across tiers (within ±2
percentage points of the cohort overall rate of 28.1\% on PI-CAI and
28.6\% on Prostate158); a χ² test of independence between tier and
suppressed/residual status was non-significant for both cohorts.
Negative-case FPs are enumerated separately because no case-specific
cancer vector exists for those cases. Tertile boundaries: PI-CAI p33 =
0.151, p67 = 0.764; Prostate158 p33 = 0.069, p67 = 0.816.

\end{document}